\documentclass[12pt,epsf]{article}

\usepackage{graphicx}
\usepackage{indentfirst}
\usepackage{mathrsfs}
\usepackage{amssymb}
\usepackage{amsmath}

\newcommand{\be}{\begin{equation}}

\newcommand{\ee}{\end{equation}}

\newcommand{\ba}{\begin{array}}

\newcommand{\ea}{\end{array}}

\begin{document}
\begin{titlepage}
\vspace{.5in}
\begin{flushright}
CQUeST-2008-0230
\end{flushright}
\vspace{0.5cm}

\begin{center}
{\Large\bf The vacuum bubbles in de Sitter background and black hole pair creation}\\
\vspace{.4in}

   {$\rm{Bum-Hoon \,\, Lee}$}\footnote{\it
   email:bhl@sogang.ac.kr}\,\, \rm{and}
   {$\rm{Wonwoo \,\, Lee}$}\footnote{\it email:warrior@sogang.ac.kr}\\

  {\small \it Department of Physics and BK21 Division, and Center for Quantum Spacetime, Sogang University, Seoul 121-742,
  Korea}\\

\vspace{.4in}

\vspace{.5in}
\end{center}
\begin{center}
{\large\bf Abstract}
\end{center}
\begin{center}
\begin{minipage}{4.75in}

{\small \,\,\,\, We study the possible types of the nucleation of
vacuum bubbles. We classify vacuum bubbles in de Sitter background
and present some numerical solutions. The thin-wall approximation
is employed to obtain the nucleation rate and the radius of vacuum
bubbles. With careful analysis we confirm that Parke's formula is
also applicable to the large true vacuum bubbles. The nucleation
of the false vacuum bubble in de Sitter background is also
evaluated. The tunneling process in the potential with degenerate
vacua is analyzed as the limiting cases of the large true vacuum
bubble and false vacuum bubble.  Next, we consider the pair
creation of black holes in the background of bubble solutions. We
obtain static bubble wall solutions of junction equation with
black hole pair. The masses of created black holes are uniquely
determined by the cosmological constant and surface tension on the
wall. Finally, we obtain the rate of pair creation of black
holes.}

PACS numbers: 98.80.Cq, 04.70.Dy, 11.27.+d

\end{minipage}
\end{center}
\end{titlepage}

\newpage
\section{ Introduction \label{sec1}}

The aims of this paper are to obtain the nucleation rate and the
radius of vacuum bubbles and to study the pair creation of black
holes in the presence of a bubble wall. The processes of bubble
nucleations and black hole productions may mimic Wheeler's
spacetime foam structure \cite{wheeler01} in the very early
universe, which represents the spacetime no longer smooth at the
Planck scale. Bubble nucleation \cite{voloshin, col002, bnu02,
bnu01} and Hawking-Moss type transition \cite{hawking} may occur
in the early universe. They may play an important role in
selecting out our universe or inflation \cite{guth0}. In addition,
the creation of black holes may also play an important role, as a
topology changing process, in the spacetime foam structure. Some
bubbles may have black holes. Some of them may be nucleated in the
presence of a black hole, which is studied by Hiscock \cite{his04}
and Berezin {\it et al.}\ \cite{bere001}, where the black hole act
as an effective nucleation center for a bubble formation. Others
may cause pair creation of black holes. On the other hand, the
string theory landscape paradigm has many stable and metastable
vacua \cite{landscape01}. So the tunneling process becomes a
remarkable event in this framework as well as eternal inflation
\cite{eternal01}.

We would like to understand the mechanism how the complicated
spacetime structure could be created and tunneling process occur.
Can the theory tell us which bubble is our universe if we live in
one? Why the particular bubble is favored? These questions are
related to the cosmological constant problem and anthropic
constraints. True vacuum bubbles can be nucleated within false
vacuum as well as false vacuum bubbles can be nucleated within
true vacuum. Black hole can also be created in these situation.
These events cause the complicated structure of the early
universe. Moreover, we may live in one bubble today. Our study
show that the important thing is the cosmological constant. We
will discuss that later in more detail. We want to classify all
possible types of bubbles. It is related to the cosmological
constant in our framework.

In this paper, we directly compute the nucleation rate of possible
vacuum bubbles using thin-wall approximation. We will classify the
true vacuum bubble in three classes : ``small'' bubble - ``large''
background, ``half'' bubble - ``half'' background, and ``large''
bubble - ``small'' background, the terminology will be defined
later. In Ref.\ \cite{parke}, Parke evaluated the nucleation rate
and the radius of a bubble with an arbitrary cosmological
constant. In our terminology, the bubble in Ref.\ \cite{parke}
corresponds to a ``small'' true vacuum bubble within a ``large''
false vacuum background. We carefully examine his formula to the
cases of large true vacuum bubbles, degenerate case, and false
vacuum bubbles while restricting the background as de Sitter.
Another method to account for the initial quantum field state as
the conformal vacuum with gravity was proposed \cite{rubakov01}.
The authors studied the Hawking-Moss transition as a limit of
constrained instantons and Lee-Weinberg transition \cite{lee01} as
a false vacuum bubble nucleation. The vacuum tunneling with a DBI
action is considered in Coleman-de Luccia type \cite{sarangi01}
and Hawking-Moss type transition \cite{wohns01}. A new method to
calculate the tunneling wave function that describes vacuum decay
was studied by Gen and Sasaki \cite{gsa02}. The tunneling using
junction condition and shell-mediated tunneling was studied
\cite{anso02}. The effect of the Gauss-Bonnet term on vacuum decay
was also studied \cite{cai001}. The mechanism of the nucleation of
a false vacuum bubble due to nonminimal coupling was obtained
\cite{wlee01}. The dynamics of a false vacuum bubble with an
effective negative tension due to the coupling was also studied in
Re.\ \cite{wlee02} (for a breathing bubble, see Ref.\ \cite{eigs}
and dynamics of a vacuum bubble, see Ref.\ \cite{bhle}). It was
proposed that the observed accelerated expansion of the universe
is driven by the false vacuum energy of a colored scalar field in
Ref.\ \cite{sto01}). Recently, Marvel and Wesley studied thin-wall
instanton with negative tension wall and its relation to Witten's
bubble of nothing \cite{wesley01}. The negative boundary tension
brane was analyzed in Ref.\ \cite{lehners01}. In Ref.\
\cite{marvel01}, the authors discussed the possible instanton
solutions in Euclidean de Sitter space. They studied tunneling
solution with the action having only Einstein-Hilbert term and
cosmological term without scalar field. Brown and Weinberg studied
thermal derivation of the Coleman-De Luccia tunneling \cite{bw01}.
The authors clarified the meaning of the bounce solution itself as
well as oscillating bounce solutions \cite{hackworth}.
Hawking-Moss transition was analyzed as a thermal fluctuation
\cite{wein05}. Recently, the relation between Coleman-de Luccia
and Hawking-Moss transition was studied in Ref.\ \cite{hwg01}.

Next, we will consider black hole pair creation. The vacuum of a
strong field can decay due to the Schwinger mechanism
\cite{schwin01} for particle production. Particle production is
one of the decaying methods of the background field or given
vacuum energy (for recent works, see Ref.\ \cite{spkim}). For the
black hole creation, Gross, Perry, and Yaffe \cite{yaffe} obtained
the nucleation rate for black holes in a thermal bath. The
nucleation of black holes in the de Sitter universe with a
semiclassical instability was considered by Ginsparg and Perry
\cite{per01}. The pair creation of black holes was originally
obtained by Gibbons \cite{gibb02}. Wu studied a single black hole
creation at the birth of the universe using the no-boundary
proposal \cite{zcwu} (for other frameworks, see Ref.\
\cite{khlop}). In Ref.\ \cite{ghtw}, the authors exhibited the new
decaying process of the cosmological constant by thermal
production of membranes. Eventually, the membrane collapses into a
black hole. Black hole pair creation in the early universe was
extensively investigated \cite{bousso, bou04} and in other
frameworks \cite{gaun01, mr03, wu05, lemos03}. Caldwell, Chamblin,
and Gibbons studied black hole pair creation in the presence of a
domain wall \cite{cal02} using the cut-and-paste procedure, where
the background has vanishing cosmological constant and the
probability was obtained. The repulsive property of the domain
wall give rise to black hole pair creation. The calculating method
in their work will be used in our paper. However, our solutions
give rise to the background for the black hole pair creation with
a bubble wall more naturally.

The outline of this paper is as follows: In Sec.\ 2 we investigate
the possible types of the nucleation of vacuum bubbles. We
classify true vacuum bubbles in de Sitter background into nine
types depending on whether the bubble and the background size are
small, half, or large. We also present some numerical solutions
corresponding to large bubbles with small background. We compute
analytically the nucleation rate and the radius of false vacuum
bubbles using the thin-wall approximation. We extend the analysis
to the large true vacuum bubbles and show that the formula in
Ref.\ \cite{parke} still holds in this case. Our approach is
different from those in Refs.\ \cite{marvel01, mal01} without
scalar field. We also consider more variety of types. The relation
to the Hawking-Moss transition will be mentioned. Furthermore, we
explore the case of degenerate vacua in de Sitter space. The
numerical solution of this tunneling was obtained in Ref.\
\cite{hackworth}. This case corresponds to a special transitions
of the nucleation of a true vacuum bubble and a false vacuum
bubble. We compare this transition with the formation of a domain
wall. The transition rate and the radius of the bubble are
presented. The deformed bulk shape causes the contribution of bulk
part in the action. In Sec.\ 3 we present the solution of pair
creation of black holes in the presence of a bubble wall. Some
possible types in previous section are adopted as the background
space for pair creation of black holes. We obtain static bubble
wall solutions of junction equation with black hole pair. The mass
of created black holes is uniquely determined by given
cosmological constant and surface tension on the wall. Finally, we
obtain the rate of pair creation of black holes. We summarize and
discuss our results in Sec.\ 4.

\section{The nucleation rate and the radius of vacuum
bubbles in de Sitter space \label{sec2}}

Let us consider the action
\begin{equation}
S= \int_{\mathcal M} \sqrt{-g} d^4 x \left[ \frac{R}{2\kappa}
-\frac{1}{2}{\nabla^\alpha}\Phi {\nabla_\alpha}\Phi
-U(\Phi)\right] + \oint_{\partial \mathcal M} \sqrt{-h} d^3 x
\frac{K}{\kappa}, \label{f-action}
\end{equation}
where $\kappa \equiv 8\pi G$, $g\equiv det g_{\mu\nu}$, and the
second term on the right-hand side is the boundary term
\cite{ygh}. $U(\Phi)$ is the scalar field potential with two
non-degenerate minima with lower minima at $\Phi_T$ and higher
minima at $\Phi_F$, $R$ denotes the Ricci curvature of spacetime
${\mathcal M}$, and $K$ is the trace of the extrinsic curvature of
the boundary $\partial \mathcal M$. Here we adopt the notations
and sign conventions of Misner, Thorne, and Wheeler \cite{misner}.

The bubble nucleation rate or the decay rate of background vacuum
is semiclassically given by $\Gamma/V = A e^{-B/\hbar}$, where $B$
is the difference between Euclidean action corresponding to bubble
solution and that of the background and the prefactor $A$ is
discussed in Ref.\ \cite{prefactor}, that with some gravitational
corrections in Ref.\ \cite{meta01}. We are interested in finding
the coefficient $B$.

We will take $O(4)$ symmetry for both $\Phi$ and the spacetime
metric $g_{\mu\nu}$, expecting its dominant contribution
\cite{col3}. The general $O(4)$-symmetric Euclidean metric is
written by
\begin{equation}
ds^2 = d\eta^2 + \rho^2(\eta)[d\chi^2 +
\sin^2\chi(d\theta^2+\sin^2\theta d\phi^2)].  \label{gemetric}
\end{equation}

The Euclidean field equations for $\Phi$ and $\rho$ have the form
\begin{eqnarray}
\Phi'' &+& \frac{3\rho'}{\rho}\Phi'=\frac{dU}{d\Phi}, \label{ephi} \\
\rho'' &=& - \frac{\kappa}{3}\rho (\Phi'^2 +U),
\end{eqnarray}
respectively and the Hamiltonian constraint is given by
\begin{equation}
\rho'^2 - 1 -
\frac{\kappa\rho^2}{3}\left(\frac{1}{2}\Phi'^{2}-U\right) = 0,
\label{erho}
\end{equation}
where the prime denotes the differentiation with respect to
$\eta$. We will consider only the case with initial de Sitter
space. The boundary conditions for the bounce are
\begin{equation}
\frac{d\Phi}{d\eta}\Big|_{\eta=0}=0, \,\,\,\,
\frac{d\Phi}{d\eta}\Big|_{\eta=\eta_{max}}=0, \,\,\,\,
\rho_{\eta=0}=0 , \,\,\,\, {\rm and}\,\, \rho_{\eta=\eta_{max}}=0,
\end{equation}
where $\eta_{max}$ is a finite value in Euclidean de Sitter space.
The asymptotic value of $\Phi(\eta)$ is given by $\lim_{\eta
\rightarrow \eta_{max}} \Phi(\eta) \simeq \Phi_{F/T}$, where
$\Phi_{F}$ is for a true vacuum bubble nucleation and $\Phi_{T}$
is for a false one. Because of the finiteness of $\eta_{max}$,
$\Phi(\eta_{max})$ is exponentially approaching to but not
reaching $\Phi_{F/T}$.

Now, we assume the thin-wall approximation scheme to evaluate $B$.
The validity of that in the case of a true vacuum bubble has been
examined \cite{samuel}. The Euclidean action can be divided into
three parts: $B= B_{in} + B_{wall} + B_{out}$. The configuration
of the outside of the wall will not be changed before and after
bubble formation. Thus, $B_{out}= 0$. So we only need to consider
the contribution of the wall and inside part. On the wall
\cite{bnu02}, $B_{wall}=2\pi^2 {\bar\rho}^3S_o$, where $S_o$ is
the tension of the wall. The action has a stationary point, which
determines the radius $\bar\rho$ of the bubble. The contribution
from the inside part $B_{in}$ will be computed in the following
subsections and will depend on whether it is true, false, or
degenerate vacua.

\subsection{True vacuum bubbles  \label{sec2-1}}

\begin{figure}
\begin{center}
\includegraphics[width=2.in]{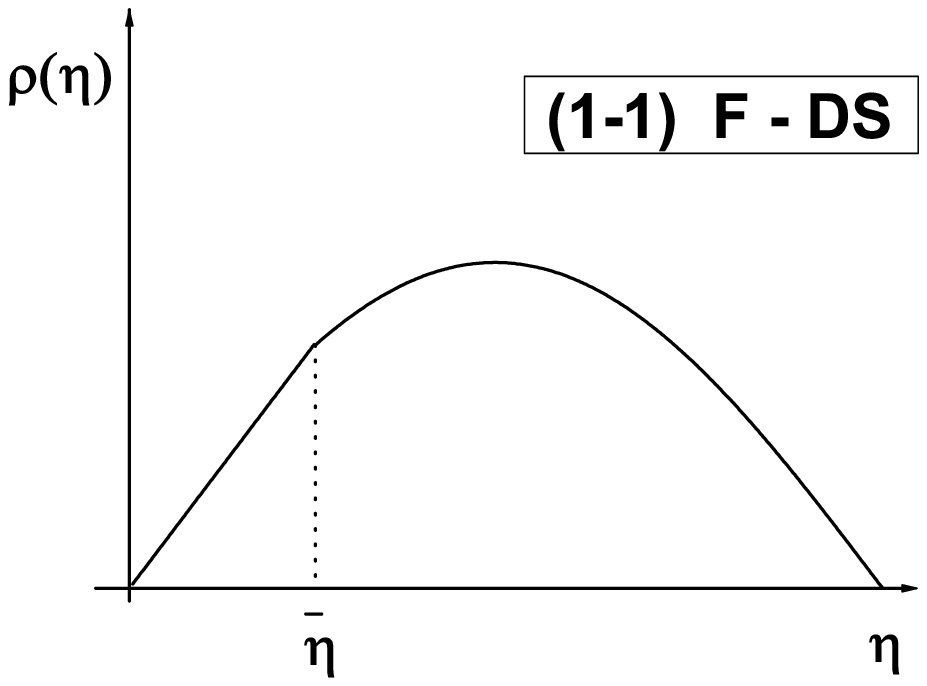}
\includegraphics[width=2.in]{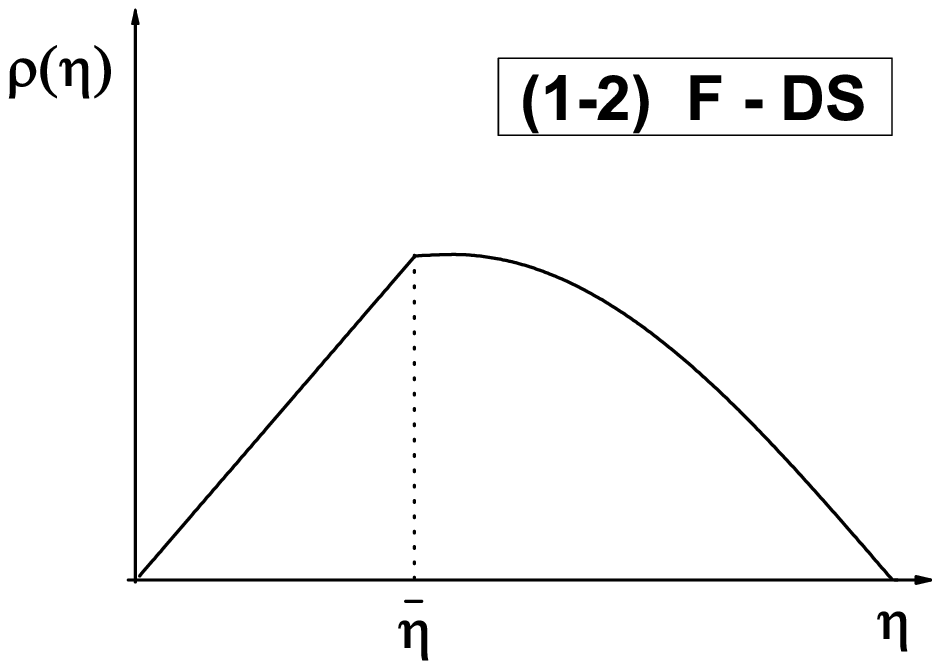}
\includegraphics[width=2.in]{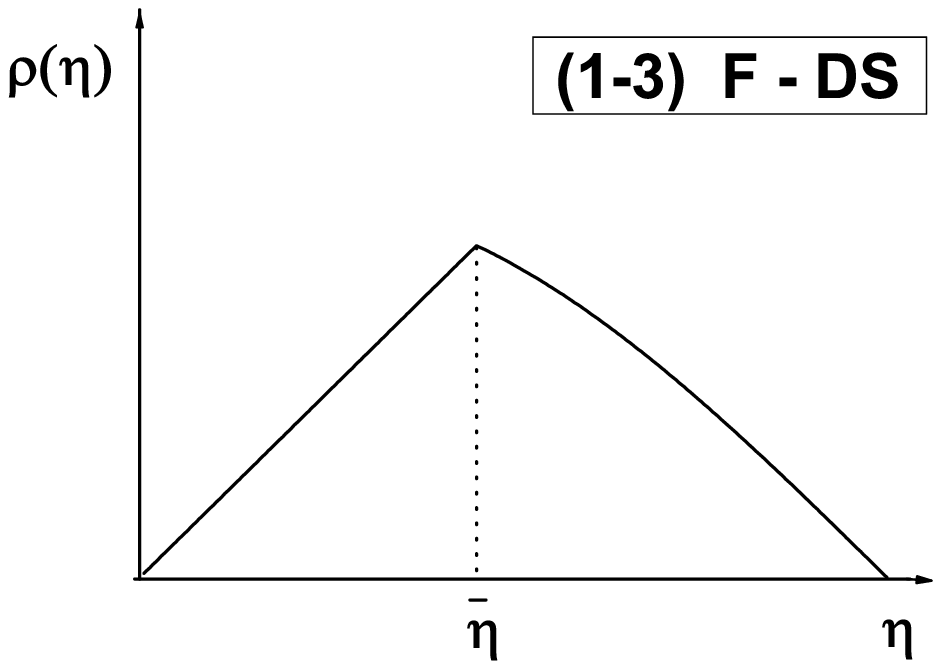}\\
\includegraphics[width=2.in]{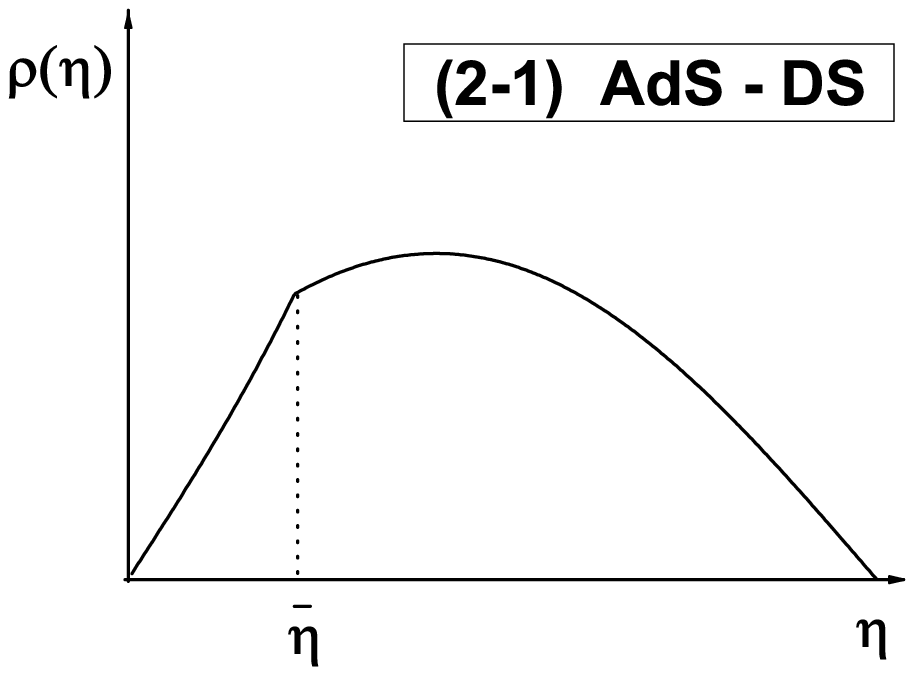}
\includegraphics[width=2.in]{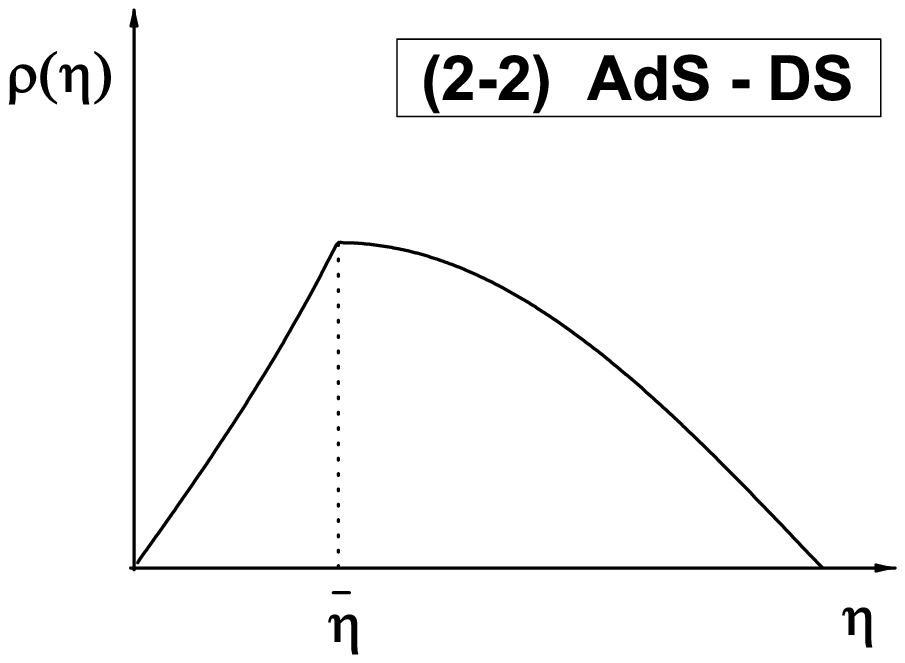}
\includegraphics[width=2.in]{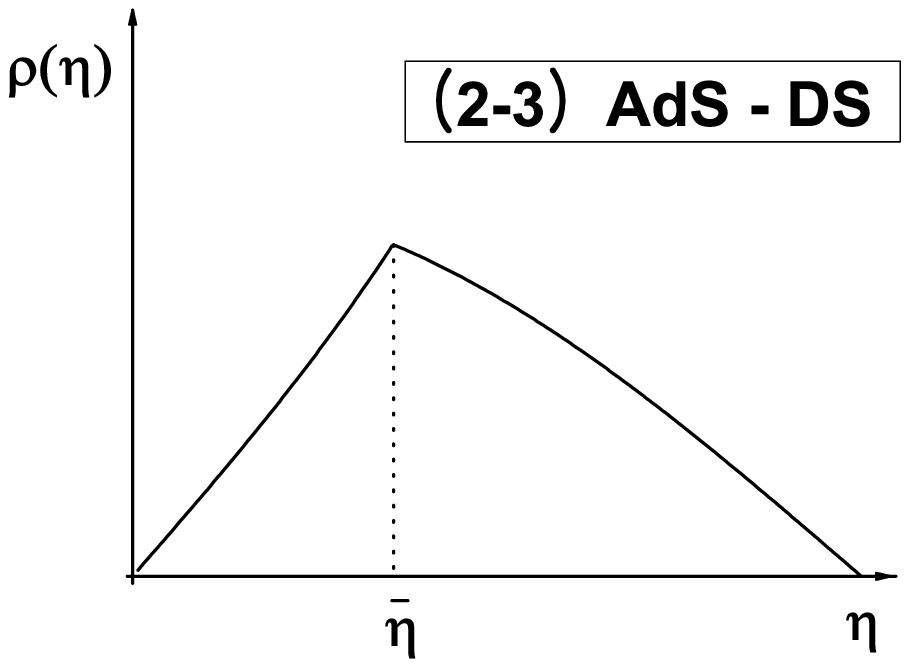}\\
\includegraphics[width=2.in]{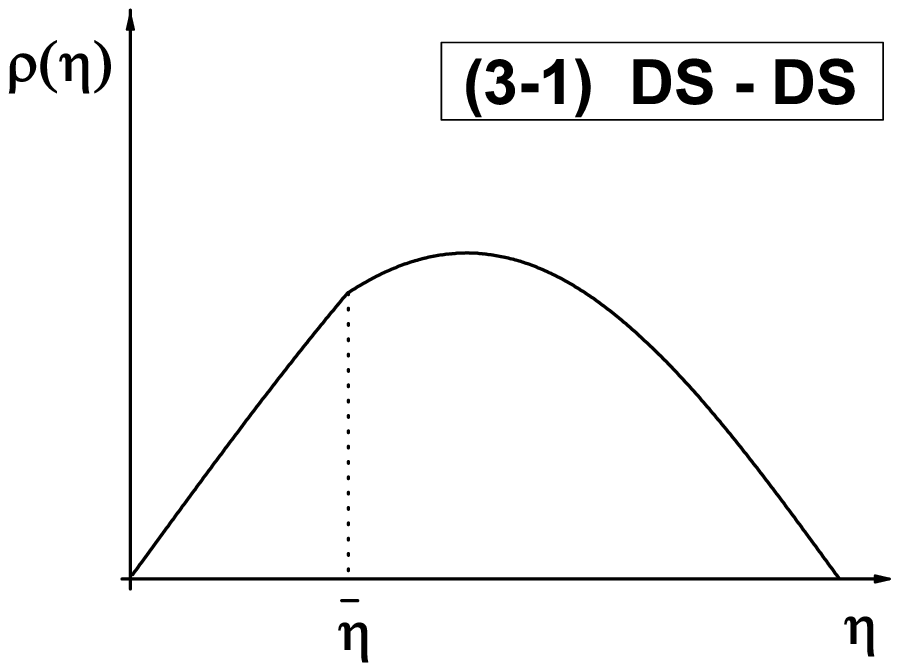}
\includegraphics[width=2.in]{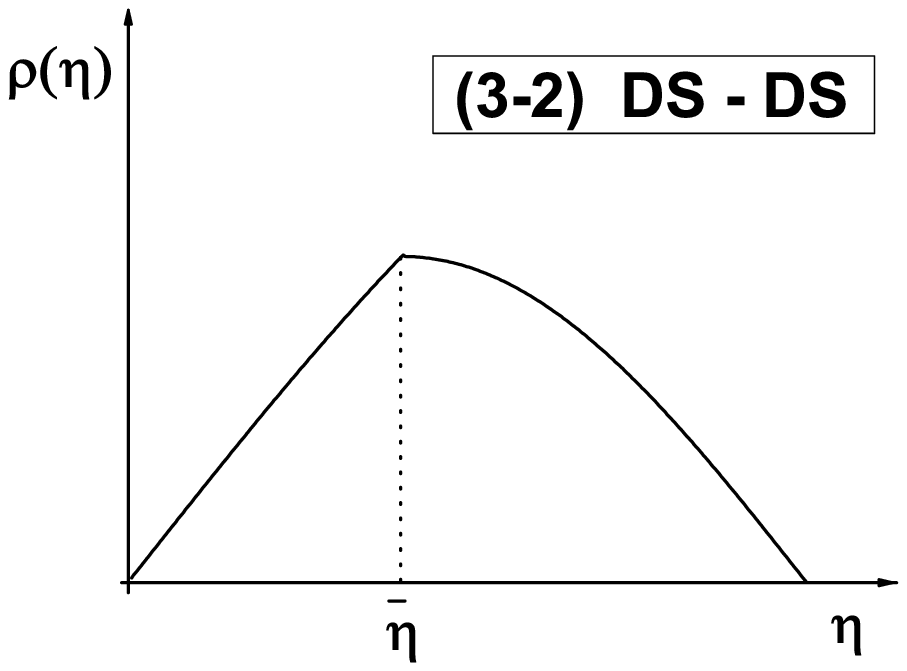}
\includegraphics[width=2.in]{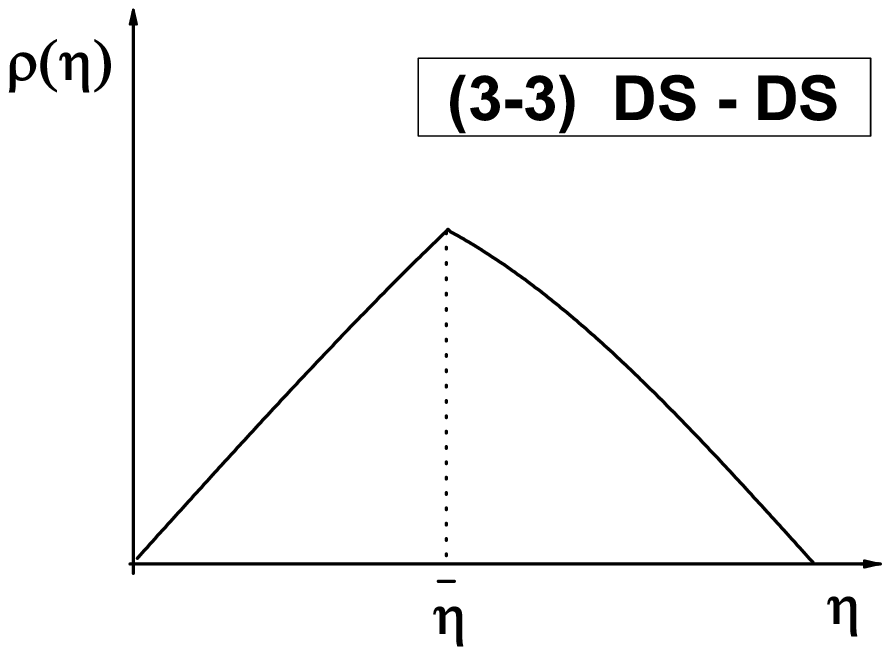}\\
{\bf Figure 1-A}
\end{center}
\begin{center}
\includegraphics[width=2.in]{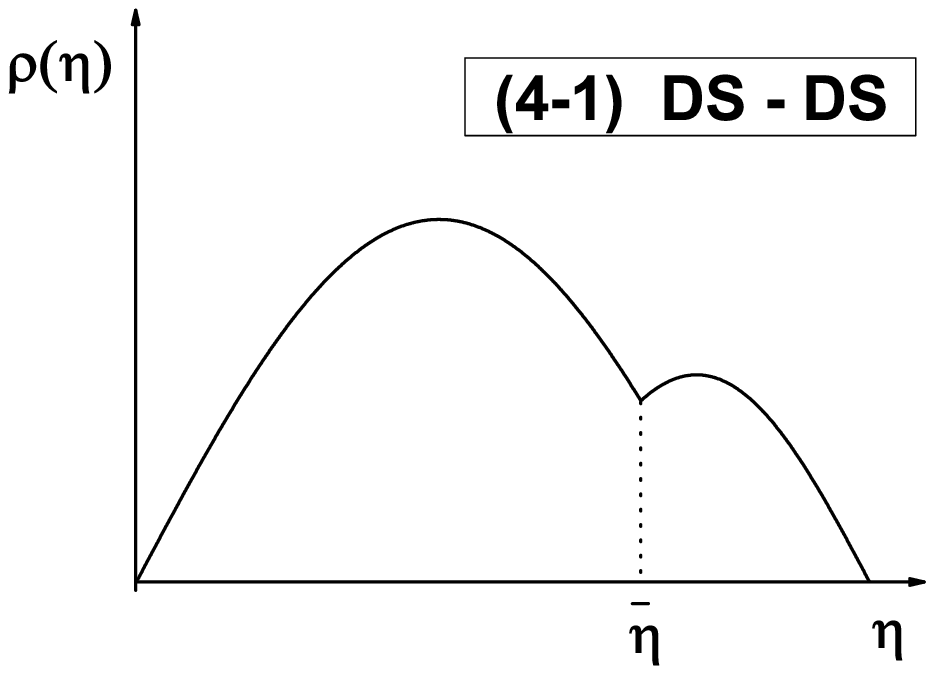}
\includegraphics[width=2.in]{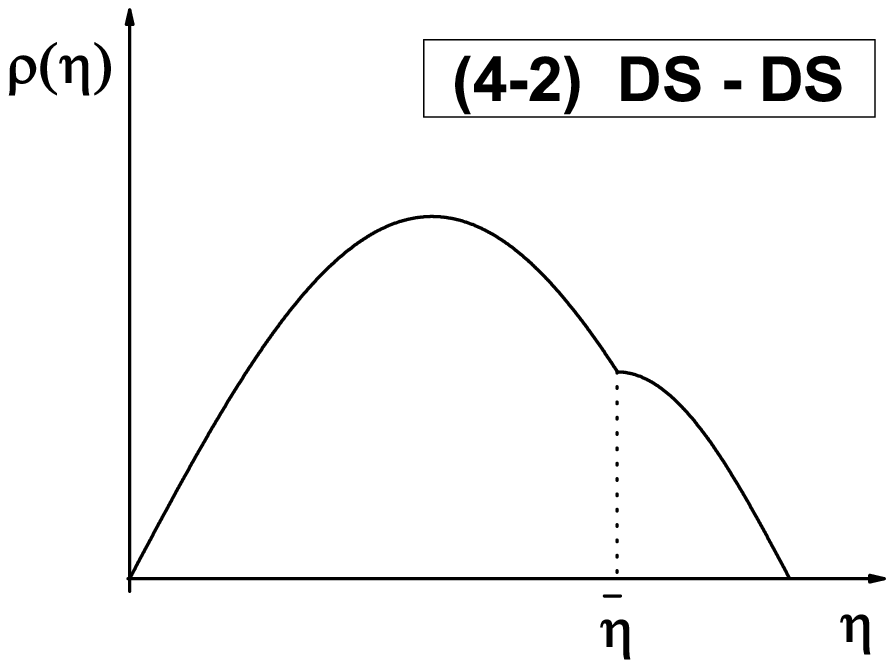}
\includegraphics[width=2.in]{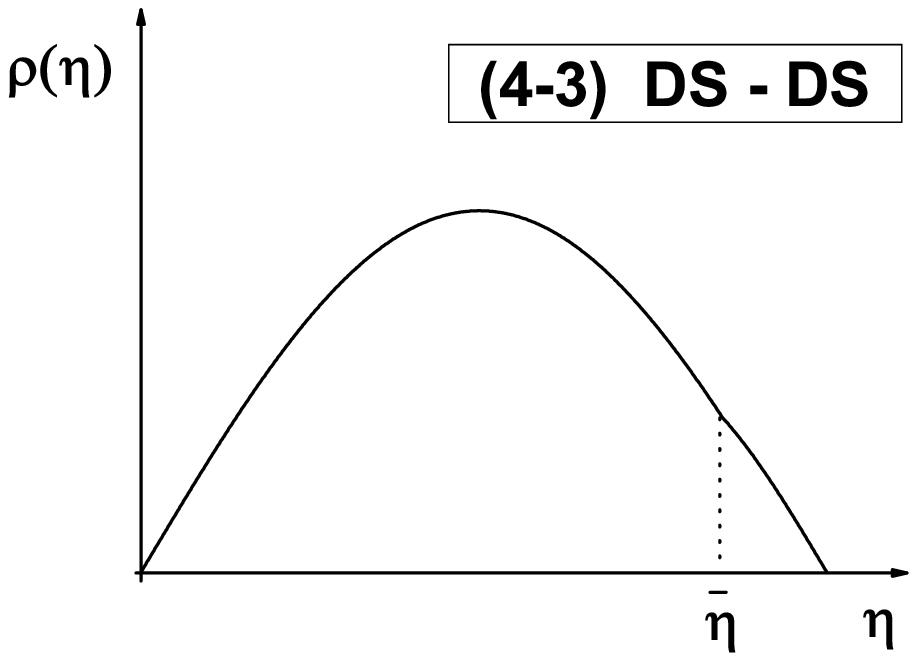}\\
{\bf Figure 1-B}
\end{center}
\caption{\footnotesize{The schematic diagram for 12 possible types
of true vacuum bubbles or matching with the thin-wall
approximation. The $\bar{\eta}$ indicates the location of the
wall. All the nine cases in Figure 1-A are possible solutions. The
cases (4-1) - (4-3) in Figure 1-B don't have the stationary point
in the action, allowing no solutions.}} \label{fig:fig01}
\end{figure}

In this section we consider possible types of true vacuum bubbles.
We consider the only $U_F>0$ so that the exterior geometry of the
bubble will remain to be de Sitter (dS) space. The true vacuum
bubble can be classified according to the interior geometry and
the size of the bubble. The interior geometry of a true vacuum
bubble can be different depending on $U_T=0$(flat), $U_T<0$(AdS),
and $U_T>0$(dS). The schematic diagrams of twelve possible types
of solutions with different shapes are illustrated in Fig.\
\ref{fig:fig01}: The figures of the first row (1-1) - (1-3) in
Fig.\ \ref{fig:fig01}-A represent flat interior geometry, those of
the second row (2-1) - (2-3) represent AdS interior geometry, and
those of the third (3-1) - (3-3) and the fourth row (4-1) - (4-3)
in Fig.\ \ref{fig:fig01}-B represent dS interior geometry,
respectively.

The cases in Fig.\ \ref{fig:fig01}-B don't have stationary point
in the action. Hence these solutions are not possible. The result
is consistent with that of Ref.\ \cite{marvel01}. The authors
analyzed the solution using the Israel matching condition
\cite{isr01}.

The de Sitter background in Fig.\ \ref{fig:fig01}-A can be
naturally called ``small'' for the first column, ``half'' for the
second column, and ``large'' for the third column. As for the true
vacuum bubbles we will call the bubble to be ``small'', ``half'',
and ``large'' if the de Sitter background is called ``large'',
``half'', and ``small'', respectively. With this terminology, the
three diagrams (1-1), (2-1), and (3-1) in the first column
correspond to the small bubble or large background, those in the
second column correspond to half sized bubble or background, and
the three diagrams (1-3), (2-3), and (3-3) in the third column
correspond to the large bubble or small background.

The small bubble cases of (1-1), (2-1), and (3-1) were obtained in
Refs.\ \cite{bnu02, parke}. The cases of (1-2), (1-3), (3-3),
(4-1), and (4-3) are considered in Refs.\ \cite{marvel01, mal01},
with different action from ours, as a mechanism of reducing the
cosmological constant \cite{bro01}. There is no scalar field in
their action and hence the junction condition was imposed on the
bubble wall. In those papers, the magnitude of bubble wall's
tension determines the type of solution. But we keep the tension
constant in this paper.

We now compute the contribution from inside the wall $B_{in}$. The
expression depends on the size of the bubble and the geometry. For
the small bubble, it is given by \cite{parke}
\begin{eqnarray}
B_{in} &=& 4\pi^2 \int^{\bar\rho}_{0} d\rho \left[ \frac{\rho^3U_T
- \frac{3\rho}{\kappa}}{(1-\frac{\kappa}{3}\rho^2
U_T)^{1/2}}-\frac{\rho^3U_F -
\frac{3\rho}{\kappa}}{(1-\frac{\kappa}{3}\rho^2 U_F)^{1/2}}
\right] \nonumber \\
&=& \frac{12\pi^2}{\kappa^2} \left[
\frac{(1-\frac{\kappa}{3}U_T\bar\rho^2)^{3/2}-1}{U_T} -
\frac{(1-\frac{\kappa}{3}U_F\bar\rho^2)^{3/2}-1}{U_F} \right].
\label{bin010}
\end{eqnarray}
The formula will be used for the first and second column in Fig.\
\ref{fig:fig01}.

For the general expression corresponding to the third column for
large bubbles and the fourth row in Fig.\ \ref{fig:fig01}, it is
important to divide the integration range separately. If the dS
bubble size is larger than half of its dS size as in the cases in
the fourth row, the integration range of the true vacuum region in
$B_{in}$ should be divided into two parts. Likewise, if the size
of dS background is large than the half of its dS size, the range
should be also divided into two parts. For example, $B_{in}$ in
the case (4-3) is given by
\begin{eqnarray}
B_{in} &=& 4\pi^2 \int^{\rho_{max}}_{0} d\rho \left[
\frac{\rho^3U_T - \frac{3\rho}{\kappa}}{(1-\frac{\kappa}{3}\rho^2
U_T)^{1/2}}-\frac{\rho^3U_F -
\frac{3\rho}{\kappa}}{(1-\frac{\kappa}{3}\rho^2 U_F)^{1/2}}
\right] \nonumber \\
&-& 4\pi^2 \int^{\bar\rho}_{\rho_{max}} d\rho \left[
\frac{\rho^3U_T - \frac{3\rho}{\kappa}}{(1-\frac{\kappa}{3}\rho^2
U_T)^{1/2}}-\frac{\rho^3U_F -
\frac{3\rho}{\kappa}}{(1-\frac{\kappa}{3}\rho^2 U_F)^{1/2}}
\right], \label{lee001-wein}
\end{eqnarray}
where $\rho_{max}(F/T)=\sqrt{3/\kappa U_{F/T}}$ for dS geometry.
This can be seen by the relation
\begin{equation}
d\rho =\pm d\eta \left[1-\frac{\kappa\rho^2U}{3}\right]^{1/2},
\label{lee001-inw}
\end{equation}
where $+$ for $0 \le \eta < \eta_{max}/2$, $0$ for
$\eta=\eta_{max}/2$, and $-$ for $\eta_{max}/2 < \eta \le
\eta_{max}$. It turns out that the cases (4-1) - (4-3) in Fig.\
\ref{fig:fig01} don't have the stationary point in the action,
allowing no solutions.

We now consider each column in more detail.

In the case of the first three rows in Fig.\ \ref{fig:fig01}, the
form of the nucleation rate and radius are mainly determined by
the value $\tilde{\kappa}$ \cite{wlee01}. To obtain the bubble
radius, the coefficient $B$ is demanded to have the stationary
value with respect to $\bar\rho$ variation. There are two bubbles
with the same radius formula because Euclidean de Sitter space has
the topology of $S^4$. One is the the small bubble (the cases of
(1-1), (2-1), and (3-1)) and the other is the large bubble (the
cases of (1-3), (2-3), and (3-3)) depending on the relation among
parameters. For the former, $U_F-U_T > \frac{3\kappa S_o^2}{4}$.
For the latter, $U_F-U_T < \frac{3\kappa S_o^2}{4}$. The existence
of the above two types of bubbles can also be seen numerically by
giving the different value of a numerical parameter,
$\tilde{\kappa}=\frac{\mu^2}{\lambda}\kappa$, while keeping the
constant surface tension, $\tilde{S}_o = 2/3$ \cite{wlee01}. Half
bubbles are possible under the condition $U_F-U_T = \frac{3\kappa
S_o^2}{4}$. The cases in (1-1), (2-1), and (3-1) have the small
numerical value of $\tilde{\kappa}$ than that of the cases in
(1-3), (2-3), and (3-3). For large bubble, $\tilde{\eta}$ exceeds
$\frac{\tilde{\eta}_{max}}{2}$ while the particle remains near the
true vacuum state for a long time in the inverted potential. This
can be related to the fact that $\tilde{\eta}_{max}$ becomes
smaller as $\tilde{\kappa}$ becomes larger. Other point of view of
varying the surface tension is in Refs.\ \cite{wesley01, marvel01,
mal01}. Although the bubbles in the first and third columns have
the same formula for the radius, they have different numerical
values in general.

The first column in Fig.\ \ref{fig:fig01}-A for the cases of the
cases (1-1), (2-1), and (3-1) corresponds to the small true vacuum
bubble or the large false vacuum dS background. The small bubbles
occur when $U_F-U_T>3\kappa S^2_o/4$. This condition corresponds
to $1>(\frac{\bar{\rho}_o}{2\lambda_2})^2$. The integral range
need not to be divided as in Eq.\ (\ref{bin010}) and the results
for these cases were already obtained by Parke in Ref.\
\cite{parke}
\begin{equation}
\bar\rho^2 = \frac{\bar\rho_o^2}{D}, \;\;\;
B=\frac{2B_o[\{1+(\frac{\bar{\rho}_o}{2\lambda_1})^2\} - D^{1/2}
]}{[(\frac{\bar{\rho}_o}{2\lambda_2})^4
\{(\frac{\lambda_2}{\lambda_1})^4 -1 \}D^{1/2} ]}, \label{case2-1}
\end{equation}
where
$D=\left[1+2(\frac{\bar\rho_o}{2\lambda_1})^2+(\frac{\bar\rho_o}{2\lambda_2})^4
\right]$, $\lambda_1^2=[3/\kappa(U_F+U_T)]$, and
$\lambda_2^2=[3/\kappa(U_F-U_T)]$. $\bar{\rho}_o =3S_o/\epsilon$
and $B_o=27\pi^2S_o^4/2\epsilon^3$ are the bubble radius and the
nucleation rate in the absence of gravity, respectively.
$\epsilon=U_F-U_T$. Here we used
$(1-\frac{\kappa}{3}U_F\bar\rho^2)^{3/2}
=\frac{[1-(\frac{\bar{\rho}_o}{2\lambda_2})^2]^3}{D^{3/2}}$.

Case (2-1) and case (3-1) correspond to the small AdS and dS
bubble, respectively. Case (1-1) corresponds to the small flat
bubble. The radius of the bubble and the nucleation rate can be
simplified after substituting $U_T=0$ into Eq.\ (\ref{case2-1}) as
\begin{equation}
\bar{\rho} = \frac{\bar{\rho}_o}{1+ (\bar{\rho}_o/2\lambda)^2},
\;\;\; B= \frac{B_o}{[1+ (\bar{\rho}_o/2\lambda)^2]^2},
\label{case1-1}
\end{equation}
where $\lambda^2 = 3/\kappa\epsilon$. The results was first
obtained in Ref.\ \cite{bnu02}.

The second column corresponds to the half true vacuum bubble i.e.
the half false vacuum background. These cases can be obtained from
Parke's results by taking the limit of the radius of the bubble
$\bar{\rho}$ equal to the size of the dS background,
$\bar\rho=\rho_{max}(F)$. This half bubble condition gives
$U_F-U_T=3\kappa S^2_o/4$. This condition corresponds to
$1=(\frac{\bar{\rho}_o}{2\lambda_2})^2$. Under this condition, $D$
defined below Eq.\ (\ref{case2-1}) can be simplified as $D=(S^2_o
+ \frac{4}{3\kappa}U_T)(4/S^2_o)$.

If we substitute $\bar\rho=\rho_{max}(F)$ into Parke's formula,
the radius of the bubble and the nucleation rate can be evaluated
as follows
\begin{equation}
\bar{\rho} = \frac{2}{\kappa\sqrt{S^2_o + \frac{4}{3\kappa}U_T}},
\;\;\; B=\frac{2B_o} {(\frac{\bar{\rho}_o}{2\lambda_2})^4
[(\frac{\lambda_2}{\lambda_1})^4 -1 ] D^{1/2}
}\left[\frac{8U_F}{3\kappa S^2_o} - D^{1/2} \right].
\label{case2-2}
\end{equation}

\begin{figure} [t]
\begin{center}
\includegraphics[width=2.6in]{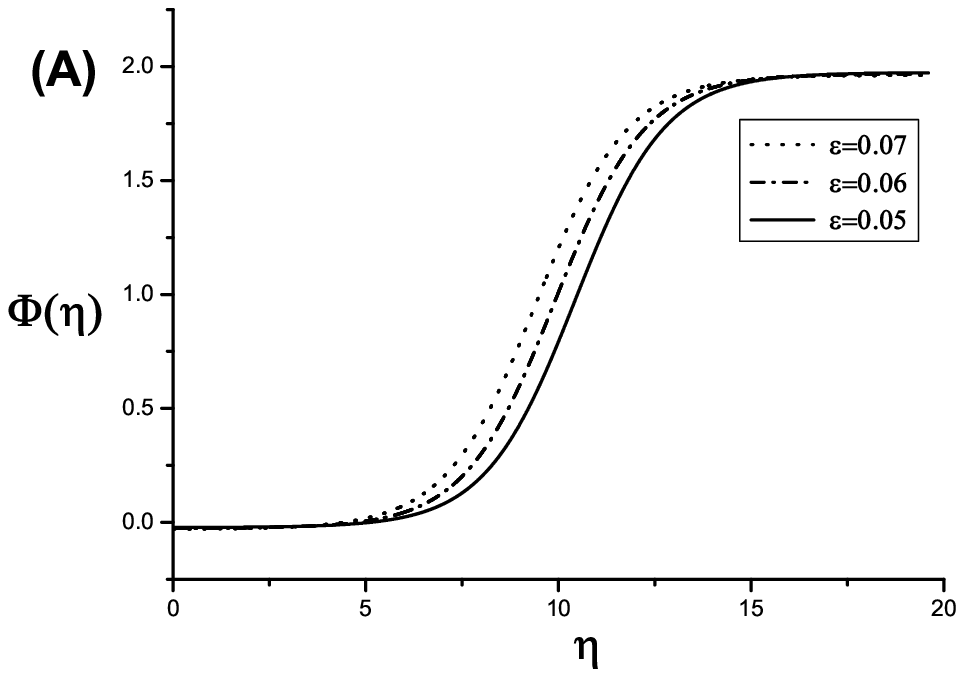}
\includegraphics[width=2.6in]{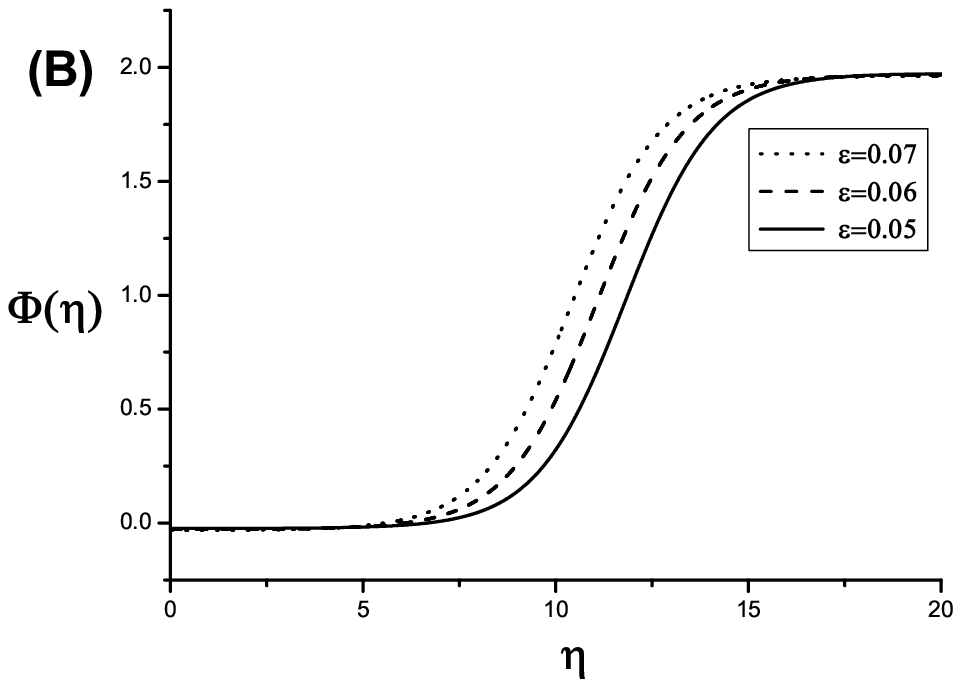}\\
\includegraphics[width=2.6in]{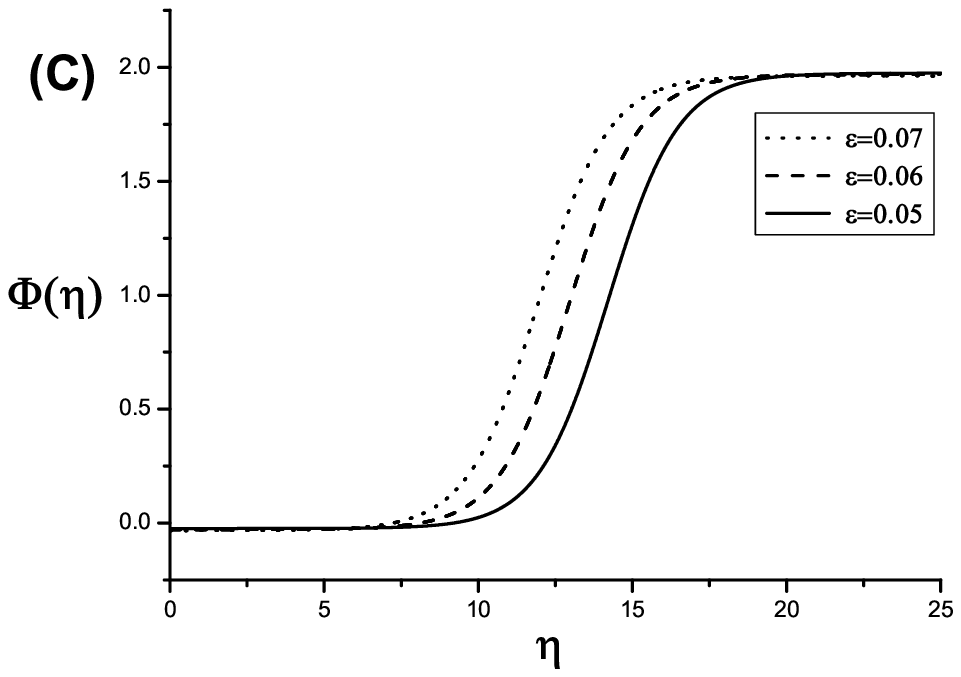}
\includegraphics[width=2.6in]{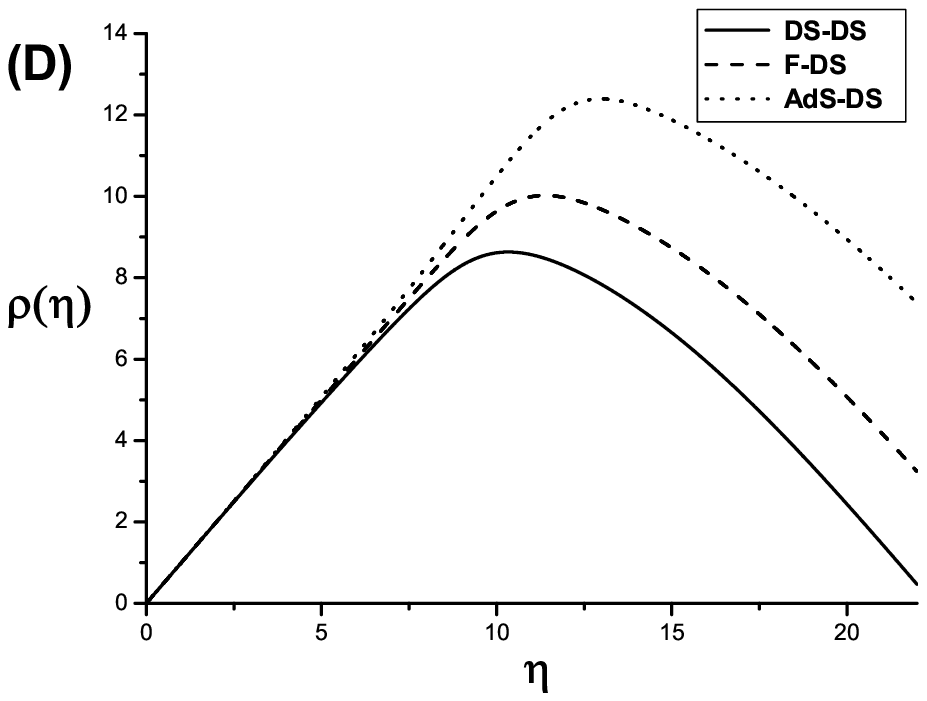}\\
\end{center}
\caption{\footnotesize{Numerical solutions for several types of
true vacuum bubbles. We take $\tilde{\kappa}\simeq0.3849$ for all
cases. The first three figures are: (A) large flat bubble - small
dS background corresponding to (1-3) in Fig.\ \ref{fig:fig01}; (B)
large AdS bubble - small dS background corresponding to (2-3); (C)
large dS bubble - small dS background corresponding to (3-3). In
Fig.\ (D), we take $\tilde{\epsilon}=0.07$. We can see from these
figures that the radius of a bubble becomes larger as
$\tilde{\epsilon}$ becomes smaller.}} \label{fig:fig02}
\end{figure}

Case (2-2) and case (3-2) correspond to the half AdS and dS
bubble, respectively. Case (1-2) corresponds to the half flat
bubble. The radius of a bubble and the nucleation rate can be
simplified after substituting $U_T=0$ into Eq.\ (\ref{case2-2}) as
\begin{equation}
\bar{\rho} = \frac{2}{S_o \kappa}, \;\;\; B=
\frac{8\pi^2}{S^2_o\kappa^3}. \label{case1-2}
\end{equation}

Finally we consider the third column of Fig.\ \ref{fig:fig01}-A
corresponding to the large true vacuum bubbles i.e. the small
false vacuum background. For the cases (1-3), (2-3), and (3-3) in
the third column in Fig.\ \ref{fig:fig01}, the integral range for
the background should be divided into two parts, hence $B_{in}$ is
then given by
\begin{eqnarray}
B_{in} &=& 4\pi^2 \left[ \int^{\bar{\rho}}_{0} \frac{\rho^3U_T -
\frac{3\rho}{\kappa}}{(1-\frac{\kappa}{3}\rho^2 U_T)^{1/2}} d\rho
 -\int^{\rho_{max}}_{0} \frac{\rho^3U_F - \frac{3\rho}{\kappa}}{(1-\frac{\kappa}{3}\rho^2
U_F)^{1/2}} d\rho + \int^{\bar{\rho}}_{\rho_{max}} \frac{\rho^3U_F
- \frac{3\rho}{\kappa}}{(1-\frac{\kappa}{3}\rho^2 U_F)^{1/2}}
d\rho \right] \nonumber \\
&=& \frac{12\pi^2}{\kappa^2} \left[
\frac{(1-\frac{\kappa}{3}U_T\bar\rho^2)^{3/2}-1}{U_T} -
\frac{(1-\frac{\kappa}{3}U_F\bar\rho^2)^{3/2}-1}{U_F} +
2\frac{(1-\frac{\kappa}{3}U_F\bar\rho^2)^{3/2}}{U_F} \right].
 \label{lee003-wein}
\end{eqnarray}
This is clearly different from Eq.\ (\ref{bin010}).

The critical bubble size $\bar{\rho}$ can be obtained by
minimizing $B=B_{in}+B_{wall}$. The large bubbles occur when
$U_F-U_T<3\kappa S^2_o/4$. This condition corresponds to
$(\frac{\bar{\rho}_o}{2\lambda_2})^2
> 1$. Case (2-3) and case (3-3) correspond to the large AdS and dS
bubble, respectively. It turns out that the nucleation rate and
the radius are given by the same form as in Eq.\ (\ref{case2-1})
even though $B_{in}$ is different. Here we use
$(1-\frac{\kappa}{3}U_F\bar\rho^2)^{3/2}
=\frac{[(\frac{\bar{\rho}_o}{2\lambda_2})^2-1]^3}{D^{3/2}}$. Case
(1-3) corresponds to the large flat bubble. The radius and the
nucleation rate of a bubble have the same form as in Eq.\
(\ref{case1-1}).

To summarize, the results of the first column in Fig.\
\ref{fig:fig01} were already obtained in Ref.\ \cite{parke} and
the results of the second column can be obtained from Eq.\
(\ref{case2-1}) using the half size condition $U_F-U_T=3\kappa
S^2_o/4$. The results of the third column have the same form as in
Eq.\ (\ref{case2-1}) even though $B_{in}$ is different. We
concluded that there is only one formula by Parke for the possible
types of the true vacuum bubble regardless of the size of the
bubble or different $B_{in}$.

The true vacuum bubbles corresponding to the first three rows in
Fig.\ \ref{fig:fig01} exist beyond the thin-wall approximation. As
some examples, we present numerical solutions in Fig.\
\ref{fig:fig02} corresponding to (1-3), (2-3), and (3-3) in the
third column of Fig.\ \ref{fig:fig01}-A.

\subsection{False vacuum bubbles \label{sec2-2}}

We now consider the nucleation of a false vacuum bubble in de
Sitter background. This was first studied by Lee and Weinberg
\cite{lee01}. We study four cases. The three cases will correspond
to the third row in Fig.\ \ref{fig:fig01} with $\eta \rightarrow
\eta_{max}-\eta$. The fourth case will be Hawking-Moss transition.

The first case corresponds to the reflected diagram of (3-1) in
Fig.\ \ref{fig:fig01}. The authors in Ref.\ \cite{lee01} have
obtained the nucleation of a false vacuum bubble within the true
vacuum background and obtained the ratio between the decay rate of
the true vacuum and that of the false vacuum. In this work, we
assume the thin-wall approximation is valid. Using the
approximation, we obtain the nucleation rate and the radius of a
false vacuum bubble. The contribution from inside the wall is
given by
\begin{eqnarray}
B_{in} &=& S^b(in)-S^T(in)= 4\pi^2 \int^{\rho_{max}}_{0} d\rho
\left[ \frac{\rho^3U_F -
\frac{3\rho}{\kappa}}{(1-\frac{\kappa}{3}\rho^2
U_F)^{1/2}}-\frac{\rho^3U_T -
\frac{3\rho}{\kappa}}{(1-\frac{\kappa}{3}\rho^2 U_T)^{1/2}}
\right] \nonumber \\
&-& 4\pi^2 \int^{\bar\rho}_{\rho_{max}} d\rho \left[
\frac{\rho^3U_F - \frac{3\rho}{\kappa}}{(1-\frac{\kappa}{3}\rho^2
U_F)^{1/2}}-\frac{\rho^3U_T -
\frac{3\rho}{\kappa}}{(1-\frac{\kappa}{3}\rho^2 U_T)^{1/2}}
\right] \nonumber \\
&=& \frac{12\pi^2}{\kappa^2} \left[
\frac{(1-\frac{\kappa}{3}U_T\bar\rho^2)^{3/2}+1}{U_T} -
\frac{(1-\frac{\kappa}{3}U_F\bar\rho^2)^{3/2}+1}{U_F} \right],
\label{lee-wein}
\end{eqnarray}
where we used $\rho^2_{max}(F/T)=\frac{3}{\kappa U_{F/T}}$. Thus
the action becomes
\begin{equation}
B = \frac{12\pi^2}{\kappa^2} \left[
\frac{(1-\frac{\kappa}{3}U_T\bar\rho^2)^{3/2}+1}{U_T} -
\frac{(1-\frac{\kappa}{3}U_F\bar\rho^2)^{3/2}+1}{U_F} \right] +
2\pi^2 \bar{\rho^3}S_o. \label{lee-wein}
\end{equation}
The tension of the wall is positive for the pure Einstein gravity
as in the case of the true vacuum bubble. This case occurs when
$U_F-U_T>3\kappa S^2_o/4$.

The formula for radius turns out to be the same as the case (2-1)
in Fig.\ \ref{fig:fig01}. We obtain the coefficient $B$ at this
extremum
\begin{equation}
B=\frac{2B_o[\{1+(\frac{\bar{\rho}_o}{2\lambda_1})^2\} + D^{1/2}
]}{[(\frac{\bar{\rho}_o}{2\lambda_2})^4
\{(\frac{\lambda_2}{\lambda_1})^4 -1 \}D^{1/2}]}. \label{fal00}
\end{equation}
Here we used $(1-\frac{\kappa}{3}U_F\bar\rho^2)^{3/2}
=\frac{[1-(\frac{\bar{\rho}_o}{2\lambda_2})^2]^3}{D^{3/2}}$. As a
result, the nucleation rate for the false vacuum bubble is given
by changing $-1$ into $+1$ in the numerator in Eq.\
(\ref{case2-1}). The ratio in Ref.\ \cite{lee01} can be obtained
from Eq.\ (\ref{fal00}) and Eq.\ (\ref{case2-1}).

The second case corresponds to the reflected diagram of (3-2) in
Fig.\ \ref{fig:fig01}. This case occurs when $U_F-U_T=3\kappa
S^2_o/4$. The radius of the bubble has the form as in (2-2) and
the nucleation rate is evaluated to be
\begin{equation}
B=\frac{2B_o} {(\frac{\bar{\rho}_o}{2\lambda_2})^4
[(\frac{\lambda_2}{\lambda_1})^4 -1 ] D^{1/2}
}\left[\frac{8U_F}{3\kappa S^2_o} + D^{1/2} \right].
\end{equation}
This result can be obtained if one substitutes
$\bar\rho=\rho_{max}(F)$ into Eq.\ (\ref{fal00}).

We now consider the third case corresponding to the reflected
diagram of (3-3) in Fig.\ \ref{fig:fig01}. This case occurs when
$U_F-U_T<3\kappa S^2_o/4$. The radius of the bubble has the form
as in (2-1) and the nucleation rate has the same form as in Eq.\
(\ref{fal00}). Here we use
$(1-\frac{\kappa}{3}U_F\bar\rho^2)^{3/2}
=\frac{[(\frac{\bar{\rho}_o}{2\lambda_2})^2-1]^3}{D^{3/2}}$.

We concluded that there is only one formula for the possible types
of false vacuum bubbles regardless of the size of the bubble or
different $B_{in}$ as Parke's formula.

The fourth case corresponds to Hawking-Moss transition. If we
change the upper limit of integral in Eq.\ (\ref{lee-wein}) into
$\bar\rho=\rho_{max}$ and eliminate the contribution of the wall
we can easily get the result obtained by Hawking and Moss. The
thin-wall approximation is not considered in this transition. It
describes the scalar field jumping simultaneously at the top of
the potential barrier. The properties of Hawking-Moss transition
were studied by many authors \cite{wein05, jst01}.

\subsection{The case of degenerate vacua \label{sec2-3}}

It is possible that the tunneling occurs {\it via} the potential
with degenerate vacua in de Sitter space. The numerical solution
of this tunneling was obtained in Ref.\ \cite{hackworth}. This
tunneling is possible due to the changing role of the second term
in Eq.\ (\ref{ephi}) from damping to accelerating during the
transition. This case corresponds to a special transitions of the
nucleation of a large true vacuum bubble and a false vacuum bubble
in dS background. The solution corresponds to the particle
starting near at one of two vacuum states under the influence of
damping term in one half of the inverted potential, climbing the
hill under the influence of accelerating term in the other half of
the potential, and reaching near at the other vacuum state. It
will be considered as the limiting case of (3-3) in Fig.\
\ref{fig:fig01}. This solution is the instanton solution rather
than the bounce solution in dS background.

Furthermore this transition is somewhat different from the
formation of a domain wall. Domain walls can form in any model
having a spontaneously broken discrete symmetry. The dynamical
solutions with a thin domain wall was obtained by Vilenkin
\cite{vilen03} and by Ipser and Sikivie \cite{ip03}. In general,
one-axis in cartesian coordinates is perpendicular to the domain
wall. Thus, $z$-axis $=$ constant hypersurfaces are isometric to
$(2+1)$-dimensional de Sitter space. The energy-momentum tensor is
localized at the domain wall surface, $z=0$. Thus, it behaves like
a delta function. The $(z, t)$-part is a $(1+1)$-dimensional
Rindler space:
\begin{equation}
ds^2 = dz^2 + (1-2\pi G\sigma |z|)^2 [-dt^2 + e^{4\pi G\sigma
t}(dx^2 + dy^2)],
\end{equation}
where $\sigma$ is the surface energy density. This metric has an
event horizon at $z=\pm 1/2\pi G\sigma$. In other coordinate
system, the metric becomes
\begin{equation}
ds^2 = \left( \frac{1}{2\pi G\sigma}\right) d\eta^2 + \eta^2
[-dt^2 + e^{4\pi G\sigma t}(dx^2 + dy^2)],
\end{equation}
where Vilenkin's solution consists of two copies of the
spacetime for $\eta < 1/2\pi G\sigma$, joined together along
the wall at $\eta=\pm 1/2\pi G\sigma$ \cite{vsh01, koli02, rbo01}.
It has the repulsive
property providing a mechanism for a tunneling process, in
particular a pair production of black holes.

In our case, the solution has $O(4)$ symmetry both inside and
outside the wall, the location of the wall is at $\eta=$ constant.
The existence of the wall doesn't affect both the inside geometry
and outside one because the process doesn't change the outside
geometry after the transition. Thus, the geometry depends
only on the vacuum energy. The spacetimes both inside and outside
of the wall are a 4-dimensional spherical Rindler type in
Lorentzian signature \cite{gerlach}
\begin{equation}
ds^2 = d\eta^2 + \rho^2(\eta)[ -dt^2 + \cosh^2t
(d\theta^2+\sin^2\theta d\phi^2)],
\end{equation}
where $\rho(\eta)=\sqrt{\frac{3}{\Lambda}}\sin
\sqrt{\frac{\Lambda}{3}}\eta$ for dS, $\rho(\eta)= \eta$ for flat,
and $\rho(\eta)=\sqrt{\frac{3}{\Lambda}}
\sinh\sqrt{\frac{\Lambda}{3}}\eta$ for AdS space.

If the thin-wall approximation is assumed the transition rate has
the form
\begin{equation}
B=  \frac{24\pi^2}{\kappa^2} \left[
\frac{(1-\frac{\kappa}{3}U_o\bar{\rho}^2)^{3/2}}{U_o}\right]
+2\pi^2 \bar{\rho}^3S_o,
\end{equation}
where $U_o$ is minima of the potential in dS space, the first term
is from the contribution of bulk part due to the deformed,
diminished, bulk shape and the second term is from the
contribution of the wall.

The nucleation rate and the critical radius of the wall are given
by
\begin{equation}
\bar{\rho} = \frac{2}{\kappa\sqrt{\frac{S^2_o}{4} +
\frac{4}{3\kappa}U_o}} \;\;\; {\rm and}\;\;\; B = \frac{12\pi^2
S_o}{\kappa^2 U_o\sqrt{\frac{S^2_o}{4} + \frac{4}{3\kappa}U_o}}.
\label{crw02}
\end{equation}
The location of the wall is smaller than $\eta_{max}/2$ because
the process deforms, or diminishes, the shape of $S^4$.

As for the consistency check, one can show that Parke's formula
for the large true vacuum bubble (3-3) and Eq.\ (\ref{fal00}) for
the reflected diagram of (3-3) become Eq.\ (\ref{crw02}) for the
degenerate case if the condition $U_F=U_T=U_o$ is substituted into
Parke's formula and Eq.\ (\ref{fal00}). In other words, nucleation
rates and radii of the large true vacuum bubble and false vacuum
bubble are continuously connected at the degenerate case as we
expected.

\section{Pair creation of black holes in the presence of a wall \label{sec3}}

\begin{figure}[t]
\begin{center}
\includegraphics[width=2.9in]{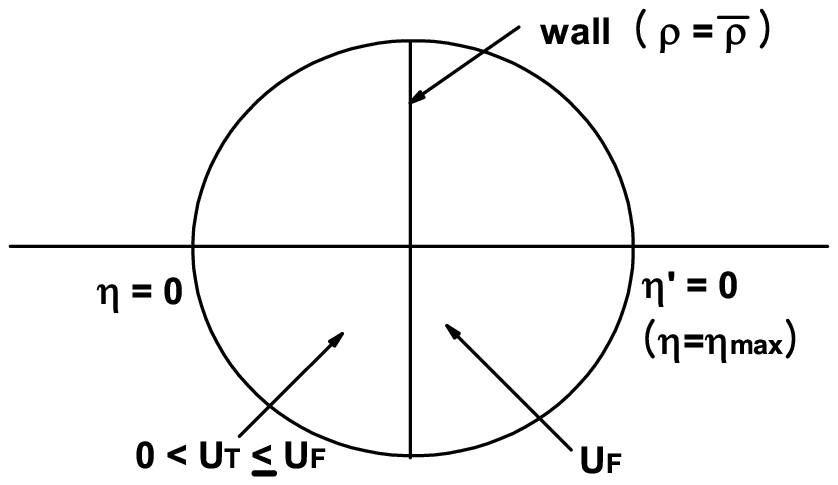}
\includegraphics[width=2.9in]{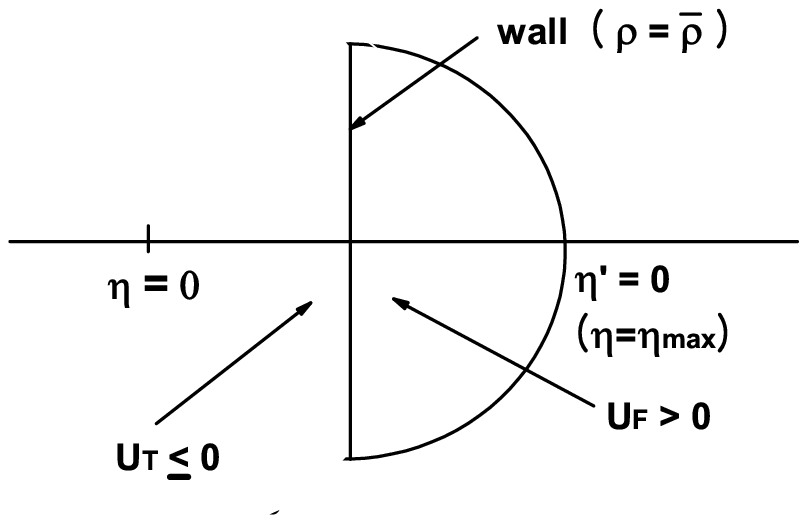}
\end{center}
\caption{\footnotesize{The schematic diagram of the bubble
geometry with the wall. The left figure corresponds to the case of
one half of the deformed de Sitter space and the other half of the
deformed de Sitter space with same or different vacuum energy. The
right figure corresponds to the case of one half of flat or
anti-de Sitter space and the other half of the deformed de Sitter
space.}} \label{fig:fig03}
\end{figure}

In this section we study the pair creation of black holes
in the background with the wall. In Ref.\ \cite{cal02}
the authors studied pair creation of black
holes in the presence of the domain wall using Euclidean junction
condition with the $Z_2$ symmetry. The authors obtained the
probability of the pair creation in the background with vanishing
cosmological constant via the cut-and-paste procedure. In this
work we will use vacuum bubbles obtained in Sec.\ \ref{sec2}. The
Fig.\ \ref{fig:fig03} shows the schematic diagram of the bubble
geometry with the wall. The geometry of outside the wall
represented as the right half in Fig.\ \ref{fig:fig03} is dS,
while inside geometry represented as the left half in Fig.\
\ref{fig:fig03} is AdS, flat, or dS depending on the cosmological
constant. We define $\eta' = \eta_{max}-\eta$ in the right half
region.

Let us consider the action
\begin{equation}
S_E= \int_{\mathcal M} \sqrt{g_E} d^4 x \left[ - \frac{R}{2\kappa}
+\frac{1}{2}{\nabla^\alpha}\Phi {\nabla_\alpha}\Phi
+U(\Phi)\right] - \oint_{\Sigma} \sqrt{h_E} d^3 x \frac{K}{\kappa}
+ \sigma \oint_{\Sigma} \sqrt{h_E} d^3 x , \label{eactwall}
\end{equation}
where the last term is a Nambu-Goto--type action on the wall.
Actually, the bulk park has two distinct spaces with boundaries
$\Sigma_+$ and $\Sigma_-$, respectively : the inside (or left) and
outside (or right) of the wall. Following the work of Chamblin and
Reall \cite{rea02}, we can obtain the Israel junction conditions
as well as the Einstein equations, and scalar field equation from
the above action. From the Einstein equations, the curvature
scalar is obtained as $R_E=\kappa[{\nabla^\alpha}\Phi
{\nabla_\alpha}\Phi +4U(\Phi)]$. From the junction conditions, the
trace of the extrinsic curvature is obtained as
$K=\frac{3}{2}\kappa\sigma$. The direction of two normal vectors
is taken to be pointing outward \cite{koli02, bl03, bou10}. In our
cases two normal vectors point in the opposite direction. After
rearranging the terms, the action becomes
\begin{equation}
S_E= - \frac{\sigma}{2} \oint_{\Sigma} \sqrt{h_E} d^3 x -
\int_{\mathcal M} \sqrt{g_E} d^4 x U(\Phi). \label{effbou}
\end{equation}
If we think over the action Eq.\ (\ref{eactwall}) as a plausible
initial setting, we have to take into account the contribution
from the relation between the boundary term and the last term.
Similar type of the action also appears in the domain wall
production. The action is different from that of usual bubble
nucleation \cite{bnu02}, which is slightly deviated from the case
in flat spacetime \cite{voloshin, col002}. In the thin wall, the
scalar field varies continuously between the true and the false
vacuum states. Thus, the boundary term doesn't contribute to the
action.

We now do the following coordinate transformations suitable for
$O(3)$ invariant configurations
\begin{eqnarray}
R&=& \sqrt{3/\Lambda} \sin (\sqrt{\Lambda/3}\eta) \sin\chi ,
\;\;\; T =  \sqrt{3/\Lambda} \tan^{-1}
\left(\tan (\sqrt{\Lambda/3}\eta) \cos\chi \right) , \;\;\; {\rm for \;\; dS} \nonumber \\
R&=& \eta\sin\chi , \;\;\; T = \eta \cos\chi , \;\;\; {\rm for \;\; flat}  \label{ct04} \\
R&=& \sqrt{3/\Lambda} \sinh (\sqrt{\Lambda/3}\eta) \sin\chi ,
\;\;\; T = \sqrt{3/\Lambda} \tanh^{-1} \left(\tanh
(\sqrt{\Lambda/3}\eta) \cos\chi \right) , \;\;\; {\rm for \;\;
AdS}.   \nonumber
\end{eqnarray}
Then the metric becomes
\begin{equation}
ds^2 = \left( 1 \pm \frac{\Lambda}{3} R^2 \right) dT^2 +
\frac{dR^2}{\left( 1 \pm \frac{\Lambda}{3} R^2 \right)} + R^2
d\Omega^2_2 , \label{o3metric}
\end{equation}
where the sign $+$ in the parentheses denotes AdS space and $-$
denotes dS space.

The geometry of dS for this work can be divided by two parts: the
inside (or left) and outside (or right) of the wall. Two halves of
the Euclidean dS spaces described by Eq.\ (\ref{o3metric}) are
joined at the wall. Each one has the range $0 \leq R <
\sqrt{3/\Lambda}$ and $0 \leq T \leq \pi \sqrt{3/\Lambda}$. We
take the de Sitter period $\beta=2\pi\sqrt{3/\Lambda}$ and
divide by $2$ into the action.

For the right half, we use $\eta'$ in Eq.\ (\ref{ct04}) instead of
$\eta$. The positions $\eta=0$ and $\eta'=0$ correspond to $R=0$.
Actually, the radii $\bar{\rho}$ of the inside wall and outside
are same regardless of the vacuum energy in the cases we will
consider. The radius $\bar{\rho}$ is related to $\bar{R}$ (the
location of the wall without black holes), as we see from Eq.\
(\ref{ct04}). The induced metric on the wall is given by
\begin{equation}
ds^2_{\Sigma} = d\tau^2 + r^2(\tau) d\Omega^2_2.
\end{equation}

We consider several types of the configurations in Fig.\
\ref{fig:fig01} as the background space for black hole pair
creation. In this work the general Euclidean junction condition
becomes
\begin{equation}
\sqrt{1-\frac{2GM}{r} \pm \frac{\Lambda_-}{3}r^2 - \dot{r}^2} +
\sqrt{1-\frac{2GM}{r}-\frac{\Lambda_+}{3}r^2 - \dot{r}^2} = 4\pi
G\sigma r,
\end{equation}
where $S_o=\sigma$ and $\cdot$ denotes the differentiation with
respect to the proper time measured by the observer moving along
with the wall. The signs $(-)$ and $(+)$ as a subscript of
$\Lambda$ represent left and right spacetime, respectively. After
squaring twice, we define the effective potential for Euclidean
junction equation to be
\begin{equation}
V_{eff}= \frac{1}{2} -\frac{1}{2} \left[ \left(2\pi G\sigma +
\frac{(\Lambda_+ \pm \Lambda_-)}{24\pi
G\sigma}\right)^2\mp\frac{\Lambda_-}{3}\right]r^2 - \frac{GM}{r},
\end{equation}
with total energy is $0$. The static bubble wall solution
satisfies the following conditions
\begin{equation}
V_{eff}(r_b) = 0 \;\;\; {\rm and} \;\;\;
\frac{dV_{eff}}{dr}\big|_{r_b} =0.
\end{equation}
The solution can exist at $r_b=3GM$. The masses of created black
holes are uniquely determined by the given cosmological constant
and surface tension on the wall:
\begin{equation}
M= \frac{1}{3G\sqrt{3\left[\left(2\pi G\sigma + \frac{(\Lambda_+
\pm \Lambda_-)}{24\pi
G\sigma}\right)^2\mp\frac{\Lambda_-}{3}\right]}}.
\end{equation}

In this process the location of the wall with black holes is given
by
\begin{equation}
r_b=3GM  \;\;\; {\rm or} \;\;\; r_b=\frac{1}{\sqrt{3\left[
\left(2\pi G\sigma + \frac{(\Lambda_+ \pm \Lambda_-)}{24\pi
G\sigma}\right)^2\mp\frac{\Lambda_-}{3}\right]}},
\end{equation}
while the location without black holes is given by
\begin{equation}
r_{wob}=\frac{1}{\sqrt{\left[ \left(2\pi G\sigma +
\frac{(\Lambda_+ \pm \Lambda_-)}{24\pi
G\sigma}\right)^2\mp\frac{\Lambda_-}{3}\right]}}.
\end{equation}

What is the effect of the black holes to the location at the wall?
Before the creation, the location is approximately at de Sitter
horizon but inside of that. After the creation, the location is
moved toward the black hole horizon.

The action for the production takes the form
\begin{equation}
S_E= S^{(B)}_E - S^{(o)}_E , \label{subt0}
\end{equation}
where $S^{(B)}_E$ and $S^{(o)}_E$ denote the action with black
holes and without black holes, respectively. Note that the
nucleation rate $S^{(B)}_E$ of the geometry with black holes is
greater than one $S^{(o)}_E$ without black holes.

{\bf (i) de Sitter with degenerate vacua}

Firstly, we consider the case of degenerate vacua in de Sitter
space. The action  is given by
\begin{eqnarray}
S^{(o)}_E &=& - \frac{\sigma}{2} \int \sqrt{h} d^3x - U_o
\int^{\tau}_{0}
d\tau \int^{\pi}_{0} d\theta \int^{2\pi}_{0} d\phi  \int^{r}_{0} dr r^2 \sin\theta \nonumber \\
&=& - \frac{\sigma}{2}\pi^2 \bar{\rho}^3_{wob} -
\frac{\pi}{2G}\sqrt{\frac{\Lambda}{3}} r^3_{wob},
\end{eqnarray}
where $\rho(\eta)=\sqrt{\frac{3}{\Lambda}}\sin
\sqrt{\frac{\Lambda}{3}}\eta$ for dS, $\rho(\eta)= \eta$ for flat,
and $\rho(\eta)=\sqrt{\frac{3}{\Lambda}}
\sinh\sqrt{\frac{\Lambda}{3}}\eta$ for AdS space, $\Lambda=8\pi G
U_o$, and the de Sitter period $\beta=2\pi\sqrt{3/\Lambda}$. We
divided the action by $2$. Our results with vanishing cosmological
constant are reduced as those in Ref.\ \cite{cal02}. We now
compare the location of a bubble wall $\bar{\rho}$ with
$r_{wob}=\frac{1}{\sqrt{\frac{\Lambda}{3} + 4\pi^2G^2\sigma^2}}$
in Eq.\ (\ref{crw02}). These two values are the same at
$\chi=\pi/2$ and $\tau=0$ (see Eq.\ (\ref{ct04})).

The action for the pair creation is evaluated to be
\begin{eqnarray}
S_{E} &=& S^C_E - \frac{\sigma}{2} \sqrt{1-\frac{2GM}{r}-
\frac{\Lambda}{3}r^2} \int^{\tau}_{0} d\tau \int^{\pi}_{0} d\theta
\int^{2\pi}_{0} d\phi r^2 \sin\theta   \nonumber \\
&-& U_o \int^{\tau}_{0}
d\tau \int^{\pi}_{0} d\theta \int^{2\pi}_{0} d\phi  \int^{r}_{0} dr r^2 \sin\theta - S^{(o)}_E \nonumber \\
&=& S^C_E - \left( 4\pi^2 G \sigma^2 r^3_b + \frac{\Lambda}{6G}
r^3_b \right) \frac{2\pi r_h\sqrt{1-(9\Lambda
G^2M^2)^{1/3}}}{|1-\Lambda
r^2_h|}  \nonumber \\
&+& \frac{\sigma}{2}\pi^2 \bar{\rho}^3_{wob} +
\frac{\pi}{2G}\sqrt{\frac{\Lambda}{3}} r^3_{wob} ,
\end{eqnarray}
where $S^C_E$ represents the contribution due to the conical
deficit angle at the black hole horizon, which has to be added in
the action $S_E$. The contribution on the conical angle was
studied in Refs.\ \cite{per01, zcwu, bou04}.

We adopt the periodicity of the time coordinate, by the
Bousso-Hawking normalization \cite{bou04} for Schwarzschild-de
Sitter (SdS) space, is $\beta=\frac{4\pi r_h\sqrt{1-(9\Lambda
G^2M^2)^{1/3}}}{|1-\Lambda r^2_h|}$ as in Ref.\ \cite{cor01}.
$r_h$ denotes $r_B$ for the black hole horizon and $r_C$ for the
cosmological horizon, respectively. In particular, this
periodicity approaches a constant value given by
$\beta_N=\frac{2\pi}{\sqrt{\Lambda}}$ in the Nariai Limit. The
statement is related to selecting of the preferred observer having
the zero acceleration. In their framework, the normalized surface
gravity is obtained and the Bousso-Hawking temperature is found.
There are recent studies on the thermodynamics of the SdS by Myung
\cite{myung01}, Nariai class by Cho and Nam \cite{cho03}, analysis
of SdS by Choudhury and Padmanabhan \cite{cp01}, other black holes
by Park \cite{par01}, and black holes in higher-order theories
\cite{garra01}.

{\bf (ii) the case (1-3) in Fig.\ \ref{fig:fig01}}

This case is that the left of the wall corresponds to flat space
and the right corresponds to de Sitter space. The action turns out
to be
\begin{eqnarray}
S_{E} &=& S^C_E - \frac{\sigma}{4} \left( \sqrt{1-\frac{2GM}{r}}
\int^{\tau_-}_{0} d\tau_- + \sqrt{1-\frac{2GM}{r}-
\frac{\Lambda_+}{3}r^2} \int^{\tau_+}_{0} d\tau_+ \right)
\int^{\pi}_{0} d\theta \int^{2\pi}_{0} d\phi r^2 \sin\theta \nonumber \\
&-& S_E(bulk) - S^{(o)}_E  \nonumber \\
&=& S^C_E - \pi\sigma r^2_b \left[ \frac{4\pi GM}{\sqrt{3}}+
\left(4\pi G\sigma r_b - \frac{1}{\sqrt{3}}\right) \frac{2\pi
r_h\sqrt{1-(9\Lambda_+
G^2M^2)^{1/3}}}{|1-\Lambda_+ r^2_h|} \right] \nonumber \\
&-& \frac{\Lambda_+}{12G} r^3_b \left(  \frac{2\pi
r_h\sqrt{1-(9\Lambda_+ G^2M^2)^{1/3}}}{|1-\Lambda_+ r^2_h|}
\right) + \frac{\sigma}{2}\pi^2 \bar{\rho}^3_{wob} +
\frac{\pi}{4G}\sqrt{\frac{\Lambda_+}{3}} r^3_{wob},
\end{eqnarray}
where the action for flat space is obtained as in \cite{cal02}.

{\bf (iii) the case (2-3) in Fig.\ \ref{fig:fig01}}

This case is that the left of the wall corresponds to anti-de
Sitter space and the right corresponds to de Sitter space. The
action is evaluated to be
\begin{eqnarray}
S_{E} &=& S^C_E -\frac{1}{4} \left( \sigma \sqrt{1-\frac{2GM}{r} +
\frac{\Lambda_-}{3}r^2} \int^{\tau_-}_{0} d\tau_-
\right.  \nonumber \\
&+& \left. \sigma \sqrt{1-\frac{2GM}{r}- \frac{\Lambda_+}{3}r^2}
\int^{\tau_+}_{0} d\tau_+ \right) \int^{\pi}_{0} d\theta
\int^{2\pi}_{0} d\phi r^2
\sin\theta - S_E(bulk) - S^{(o)}_E  \nonumber \\
&=& S^C_E - \pi\sigma r^2_b \left[
\sqrt{\frac{1}{3}+\frac{\Lambda_-}{3} r^2_b } \frac{4\pi
r_h\sqrt{\frac{1}{3}+\frac{\Lambda_-}{3}r^2_b}}{1+\Lambda_- r^2_h} \right.  \nonumber \\
 &+& \left. \left(4\pi G\sigma r_b
-\sqrt{\frac{1}{3}+\frac{\Lambda_-}{3} r^2_b } \right) \frac{2\pi
r_h\sqrt{1-(9\Lambda_+
G^2M^2)^{1/3}}}{|1-\Lambda_+ r^2_h|}  \right]  \nonumber \\
&-& \frac{r^3_b}{12G} \left[ \Lambda_+  \frac{2\pi
r_h\sqrt{1-(9\Lambda_+ G^2M^2)^{1/3}}}{|1-\Lambda_+ r^2_h|} -
\Lambda_- \frac{4\pi
r_h\sqrt{\frac{1}{3}+\frac{\Lambda_-}{3}r^2_b}}{1+\Lambda_-
r^2_h}  \right] \nonumber \\
&+& \frac{\sigma}{2}\pi^2 \bar{\rho}^3_{wob} +
\frac{\pi}{4G}\sqrt{\frac{\Lambda_+}{3}} r^3_{wob} -
\frac{\Lambda_-}{64G}\pi \bar{\rho}^4_{wob},
\end{eqnarray}
where $\beta = \frac{4\pi
r_h\sqrt{1-\frac{2GM}{r_b}+\frac{\Lambda_-}{3}r^2_b}}{1+\Lambda_-
r^2_h}$ as in Ref.\ \cite{brown04}.

{\bf (iv) the case (3-3) in Fig.\ \ref{fig:fig01}}

This case is that the left of the wall corresponds to de Sitter
space and the right corresponds to de Sitter space with different
vacua. The action turns out to be
\begin{eqnarray}
S_{E} &=& S^C_E -\frac{\sigma}{4} \left( \sqrt{1-\frac{2GM}{r} -
\frac{\Lambda_-}{3}r^2} \int^{\tau_-}_{0} d\tau_-   \right.  \nonumber \\
&+& \left. \sqrt{1-\frac{2GM}{r}- \frac{\Lambda_+}{3}r^2}
\int^{\tau_+}_{0} d\tau_+ \right) \int^{\pi}_{0} d\theta
\int^{2\pi}_{0} d\phi r^2
\sin\theta - S_E(bulk) - S^{(o)}_E  \nonumber \\
&=& S^C_E - \pi\sigma r^2_b \left[
\sqrt{\frac{1}{3}-\frac{\Lambda_-}{3} r^2_b } \frac{2\pi
r_{h_-}\sqrt{1-(9\Lambda_-
G^2M^2)^{1/3}}}{|1-\Lambda_- r^2_{h_-}|} \right.  \nonumber \\
 &+& \left. \left(4\pi G\sigma r_b
-\sqrt{\frac{1}{3}-\frac{\Lambda_-}{3} r^2_b } \right) \frac{2\pi
r_{h_+}\sqrt{1-(9\Lambda_+
G^2M^2)^{1/3}}}{|1-\Lambda_+ r^2_{h_+}|}  \right]  \nonumber \\
&-& \frac{r^3_b}{12G} \left[ \Lambda_+  \frac{2\pi
r_{h_+}\sqrt{1-(9\Lambda_+ G^2M^2)^{1/3}}}{|1-\Lambda_+
r^2_{h_+}|} + \Lambda_- \frac{2\pi
r_{h_-}\sqrt{1-(9\Lambda_- G^2M^2)^{1/3}}}{|1-\Lambda_- r^2_{h_-}|}  \right] \nonumber \\
&+& \frac{\sigma}{2}\pi^2 \bar{\rho}^3_{wob} +
\frac{\pi}{4G\sqrt{3}} r^3_{wob}(\sqrt{\Lambda_+}
+\sqrt{\Lambda_-}).
\end{eqnarray}
In this section we have obtained pair creation rate of black holes
in the four cases with different background types.

\section{Summary and Discussions}

In this paper we studied the possible types of vacuum bubbles and
calculated the radius and the nucleation rate. We have obtained
some numerical solutions as well as analytic computation using the
thin-wall approximation. We consider the only $U_F>0$ so that the
exterior geometry of the bubble will remain to be de Sitter (dS)
space. There are nine types of true vacuum bubbles, three false
vacuum bubbles, and Hawking-Moss transition in which the thin-wall
approximation is not considered. We confirmed that Parke's formula
is applicable to all types of true vacuum bubbles even though
$B_{in}$ is different. In addition, we obtained the single formula
that applies also for all types of false vacuum bubbles. There are
some conditions for classifying the bubbles. The conditions are
$U_F-U_T>3\kappa S^2_o/4$ for small bubbles, $U_F-U_T=3\kappa
S^2_o/4$ for half bubbles, and $U_F-U_T<3\kappa S^2_o/4$ for large
bubbles. The large true vacuum bubble corresponds to the case (3-3) 
and false vacuum bubble to the reflected diagram of (3-3). 
The coefficient $B$ and the radius of the former is continuously 
connected to those of the latter.

We have discussed the special case corresponding to the tunneling
of degenerate vacua in de Sitter space. This tunneling is possible
due to the changing role of the second term in Eq.\ (\ref{ephi})
from damping to acceleration during the transition. This solution
does not have any negative mode, hence corresponds to an instanton
solution rather than a bounce solution in de Sitter background. We
have compared this transition with the formation of a domain wall.
We have obtained the transition rate and the radius of a bubble.
This solution can be applied to the sine-Gordan model with
periodic potential. Moreover this solution can be applied to the
case with a complex scalar field, which has global $U(1)$
symmetry. In the potential, the transition between different
minima, by moving along the circle of minima, requires an infinite
energy because of the infinite volume of the space. The kinetic
term in the energy is proportional to the volume. Therefore, a
change in the field in the whole space requires an infinite energy
and thus the transition is impossible. However, the transition
{\it via} tunneling can be possible in de Sitter space. In
addition, this transition by moving along the circle will be
possible because of the finite volume of de Sitter space. The
process of the symmetry restoration will be also possible {\it
via} the Hawking-Moss transition. It will be interesting if this
transition can be embedded in the case with $SO(3)$ model as well
as in the cases with gauge fields in various dimensions.

We considered the pair creation of black holes in some of these
vacuum bubble backgrounds. These solutions give rise to the
background for the black hole pair creation more naturally. We
have obtained static bubble wall solutions of junction equation
with black hole pair. The masses of created black holes are
uniquely determined by given cosmological constant and the surface
energy density or tension on the wall. In this work, the masses of
black holes are same. Finally, we have obtained the rate of pair
creation of black holes by taking the difference between the
nucleation rate $S^{(B)}_E$ of the geometry with black holes and
that $S^{(o)}_E$ without black holes. In this paper we considered
black holes having only the same mass in pure Einstein gravity.
The case with different masses can be applied. Considering other
theories of gravity with the modified tension of the wall will be
also interesting.

Our solutions, even if it has simple structure, can be used to
understand the mechanism how the complicated spacetime structure
could be created in the early universe as well as tunneling
phenomena occur in the string landscape and eternal inflation.

All cases we considered have at least one de Sitter space. Thus,
we now need to argue the temperature of SdS space defined by the
surface gravity. Actually, the one side of the background with
black holes has one horizon due to a black hole because the
location of the wall for each solution is inside de Sitter
horizon. In general, SdS space has two horizons. One is black hole
event horizon and the other the cosmological horizon,
respectively. The temperature has also two types, Hawking
temperature for event horizon and Gibbons-Hawking temperature for
cosmological horizon in the standard normalization, respectively.
Thus, there was a difficulty determining the periodicity of the
time coordinate. In other words, there is a conical singularity
unless both singularity may be simultaneously removed.

Actually, the surface gravity is related to the force required at
infinity to hold a test particle hovering at the horizon. In SdS,
there is no asymptotic flat region and to select the preferred
observer with zero acceleration is difficult. Thus, one uses the
standard normalization with two different temperatures defined by
two different surface gravities. In the framework of the
Bousso-Hawking normalization, there is one point at which the
black hole attraction and the cosmological repulsion exactly
cancel out. Thus, one can define the Bousso-Hawking temperature
and the periodicity of the time coordinate.

If the background becomes SdS and we can't control the singular
properties an extra contribution from the boundary term due to
horizons, after black hole pair creation, will appear. There is no
cosmological horizon in our solution because the wall is located
within the horizon. The boundary effect due to the existence of
the wall appeared in Eq.\ (\ref{eactwall}) is already included in
our solution. The contribution due to the conical angle at the
black hole horizon has to be added in the action $S_{E}$.

\section*{Acknowledgements}

We would like to thank Kimyeong Lee, Piljin Yi, and Hyun Seok Yang
at the Korea Institute for Advanced Study for their hospitality
and valuable discussions and Yun Soo Myung, Jin-Ho Cho, Chanju
Kim, and Hyeong-Chan Kim for useful discussions and kind comments.
We would like to thank E. J. Weinberg for helpful discussions.
This work was supported by the Korea Science and Engineering
Foundation (KOSEF) grant funded by the Korea government(MEST)
through the Center for Quantum Spacetime(CQUeST) of Sogang
University with grant number R11 - 2005 - 021. W.L. was supported
by the Korea Research Foundation Grant funded by the Korean
Government(MOEHRD)(KRF-2007-355-C00014).

\end{document}